# *Ab initio* predictions for polarized deuterium-tritium thermonuclear fusion

**Guillaume Hupin**[1,2,3,a], **Sofia Quaglioni**[3] and **Petr Navrátil**[4]

[1] Institut de Physique Nucléaire, IN2P3/CNRS, Université Paris-Sud, Université Paris-Saclay, F-91406 Orsay Cedex, France
[2] CEA, DAM, DIF, F-91297 Arpajon, France
[3] Lawrence Livermore National Laboratory, P.O. Box 808, L-414, Livermore, California 94551, USA
[4] TRIUMF, Vancouver, BC V6T2A3, Canada

## Abstract

The fusion of deuterium (D) with tritium (T) is the most promising of the reactions that could power thermonuclear reactors of the future. It may lead to even more efficient energy generation if obtained in a polarized state, that is with the spin of the reactants aligned. Here, we report first-principles predictions of the polarized DT fusion using nuclear forces from effective field theory. By employing the *ab initio* no-core shell model with continuum reaction method to solve the quantum mechanical five-nucleon problem, we accurately determine the enhanced fusion rate and angular distribution of the emitted neutron and $^4$He. Our calculations demonstrate in detail the small contribution of anisotropies, placing on a firmer footing the understanding of the rate of DT fusion in a polarized plasma. In the future, analogous calculations could be used to obtain accurate values for other, more uncertain thermonuclear reaction data critical to nuclear science applications.

## Introduction

Thermonuclear reaction rates of light nuclei are critical to nuclear science applications ranging from the modeling of big-bang nucleosynthesis and the early phases of stellar burning to the exploration of nuclear fusion as a terrestrial source of energy. The low-energy regime (tens to hundreds of keV) typical of nucleosynthesis and fusion plasmas is challenging to probe due to low counting rates and the screening effect of electrons, which in a laboratory are bound to the reacting nuclei. A predictive understanding of thermonuclear reactions is therefore needed alongside experiments to achieve the accuracy and/or provide part of the nuclear data required by these applications. A salient example is the fusion of deuterium (D) with tritium ($^3$H or T) to generate a $^4$He nucleus ($\alpha$-particle), a neutron and 17.6 MeV of energy released in the form of kinetic energy of the products. This reaction, used at facilities such as ITER[1] and NIF[2] in the pursuit of sustained fusion-energy production, is characterized by a pronounced resonance at the center-of-mass (c.m.) energy of 65 keV above the free D and T nuclei due to the formation of the $J^\pi = 3/2^+$ resonance of the unbound $^5$He nucleus. Fifty years ago, it was estimated[3] that, in the ideal scenario in which the spins of the reactants are perfectly aligned in a total-spin 3/2 configuration and assuming that the reaction is isotropic, one could achieve an enhancement of the cross section by a factor of $\delta = 1.5$, thus improving the economics of fusion energy generation[4]. However, while the unpolarized cross section and some analyzing-power data exist, no correlation coefficients have been measured yet to confirm this prediction[5]. More generally, what little is known about the properties of the polarized DT fusion was inferred from measurements of the D$^3$He reaction[6].

---

[a] Correspondence and requests for materials should be addressed to G. Hupin (hupin@ipno.in2p3.fr).

The DT fusion is a primary example of a thermonuclear reaction in which the conversion of two lighter elements to a heavier one occurs through the transfer of a nucleon from the projectile (D) to the target (T). Despite the fairly small number of nucleons involved in this process, arriving at a comprehensive understanding – in terms of the laws of quantum mechanics and the underlying theory of the strong force (quantum chromodynamics) – of the interweaving of nuclear shell structure and reaction dynamics giving rise to the DT fusion already represents a formidable challenge for nuclear theory.

Towards this goal, a pioneering *ab initio* study of the DT fusion was carried out in ref. 7, using a nucleon-nucleon (NN) interaction that accurately describes two-nucleon data and representing the wave function on a basis of continuous

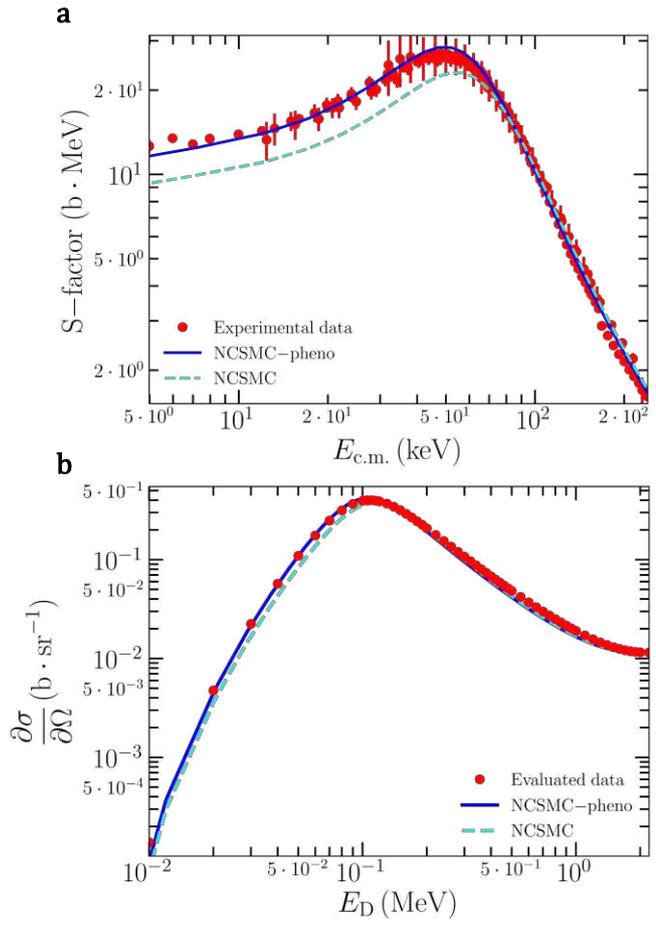

**Fig. 1 | Unpolarized DT cross sections. a** Astrophysical S-factor as a function of the energy in the center-of-mass (c.m.) frame, $E_{c.m.}$, compared to available experimental data[21-25] (with error bars indicating the associated statistical uncertainties). **b** Angular differential cross section $\left(\frac{\partial\sigma}{\partial\Omega}\right)$ as a function of the deuterium incident energy, $E_D$, at the c.m. scattering angle of $\theta_{c.m.} = 0°$ compared to the evaluated data of ref. 26. In the figures 'NCSMC' and 'NCSMC-pheno' stand for the results of the present calculations before and after a phenomenological correction of $-5$ keV to the position of the $3/2^+$ resonance.

'microscopic-cluster' states[8] made of D+T and n+$^4$He pairs in relative motion with respect to each other. However, this approach was unable to yield results of adequate fidelity, due to the omission of the three-nucleon (3N) force – disregarded for technical reasons. Numerous studies have shown that this component of the nuclear interaction is essential for the reproduction of single-particle properties[8-12], masses[13-15] and spin properties[10,16], all impactful in the present case. Besides the 3N force, the approach of ref. 7 also lacked a complete treatment of short-range five-nucleon correlations, which are crucial to arrive at the accurate description of the $3/2^+$ resonance. The formation of this rather long-lived resonance as a correlated, localized system of five nucleons built up during the fusion process is integral to the reaction mechanism. Finally, no polarization observables were calculated in the study of ref. 7.

In the following, we report on *ab initio* predictions for the polarized DT fusion using validated NN and 3N forces derived in the framework of chiral effective field theory (EFT)[17,18], a powerful tool that enables the organization of the interactions among protons and neutrons in a systematically improvable expansion linked to the fundamental theory of quantum chromodynamics. The quantum-mechanical five-nucleon problem is solved using the no-core shell model with continuum (NCSMC)[10,19], where the model space includes D+T and n+$^4$He microscopic-cluster states, plus conventional static solutions for the aggregate $^5$He system[20]. This enables a fully integrated description of the reaction in the incoming (outgoing) channel, where the reactants (products) are far apart, as well as when all five nucleons are close together. We show that this approach yields an accurate reproduction of the DT cross section for unpolarized reactants, discriminating among reaction rates from phenomenological evaluations and demonstrating in detail the small contribution of anisotropies in the vicinity of the $3/2^+$ resonance. The maximum enhancement of the polarized cross section varies as a function of the deuterium incident energy, dropping significantly above 0.8 MeV. However, such variation is slow in the narrow range of optimal energies for the reaction, resulting in a rather constant enhancement of the rate of fusion, compatible with the historic approximate estimate.

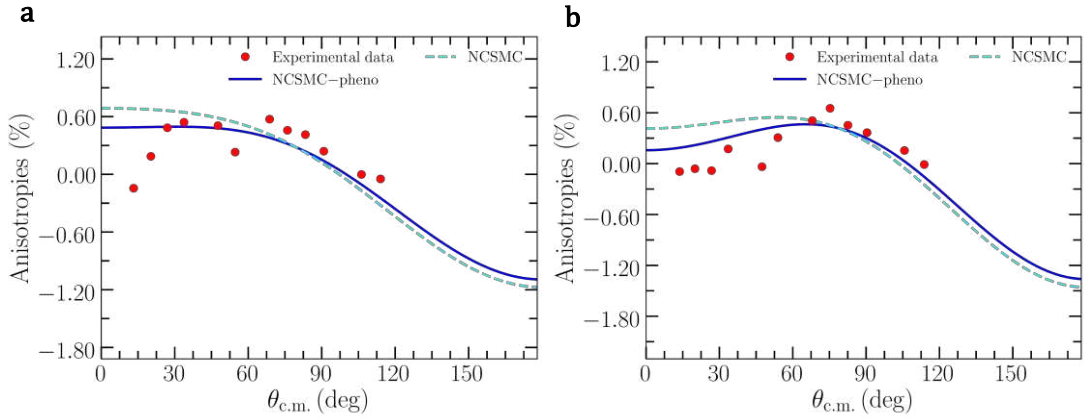

**Fig. 2 | Anisotropy in the unpolarized DT differential cross section.** Computed and measured[30] percentage of anisotropies in the unpolarized angular distribution (obtained as percentage deviations from unity of the differential cross section divided by the total – angle integrated – cross section) as a function of the scattering angle in the center-of-mass (c.m.) frame, $\theta_{c.m.}$ for two values of the deuteron incident energy, $E_D$. **a**: $E_D$ = 132.8 keV. **b**: $E_D$ = 174.7 keV. In the figures 'NCSMC' and 'NCSMC-pheno' stand for the results of the present calculations before and after a phenomenological correction of $-5$ keV to the position of the $3/2^+$ resonance and 'experimental data' are the measurements from ref. 30.

## Results

**Validation of model for unpolarized reaction observables.** We begin our study with a validation of our *ab initio* reaction method on existing experimental data for the unpolarized DT reaction[21-25]. In Fig. 1a, we review the agreement of our computed astrophysical S-factor with established measurements. The S-factor isolates the nuclear dynamics by factoring out the Coulomb component of the total reaction cross section. The experimental peak at the c.m. energy of $E_{c.m.}$ = 49.7 keV corresponds to the enhancement from the $3/2^+$ resonance of $^5$He (see Supplementary Note 2). We underpredict by 15% the experiment (green dashed line versus red circles), an outcome that can be traced back to the overestimation of the $3/2^+$ resonance centroid by a few keV, stemming from residual inaccuracies of the nuclear interaction[10]. To overcome this issue and arrive at an accurate evaluation of polarized DT reaction observables, we apply a phenomenological correction of $-5$ keV to the position of the resonance centroid, achieving a remarkable agreement with the experimental S-factor over a wide range of energies (blue line). A detailed explanation of how such correction was obtained can be found in the Methods section. As a further validation of our calculations in Fig. 1b we present the differential cross section in the center of mass frame at the scattering angle of $\theta_{c.m.} = 0°$ over a range of energies up to the deuterium breakup threshold. Our results (blue solid and green dashed lines) match the differential cross section of ref. 26 (red circles), obtained from a Legendre coefficient fit of measurements.

**Polarization enhancement and reaction anisotropy.** Having validated our calculation on precision measurements of unpolarized DT fusion, we now turn to the fusion of polarized DT fuel. The tritium has a spin of $1/2$, consequently its initial spin state is fully characterized by the Cartesian spin projection onto the axis of quantization (z axis), $q_z$. On the other hand, the deuterium is a spin-1 particle. Therefore, besides the equivalent $p_z$ projection, an extra tensor value ($p_{zz}$) is required to fully specify the spin state of the D beam. For the special case considered here, in which both reactants are aligned along the z axis, the polarized differential cross section assumes a fairly simple form and is given by,

$$\frac{\partial \sigma_{\text{pol}}}{\partial \Omega_{\text{c.m.}}}(\theta_{\text{c.m.}}) = \frac{\partial \sigma_{\text{unpol}}}{\partial \Omega_{\text{c.m.}}}(\theta_{\text{c.m.}}) \left(1 + \frac{1}{2} p_{zz} A_{zz}^{(b)}(\theta_{\text{c.m.}}) + \frac{3}{2} p_z q_z C_{z,z}(\theta_{\text{c.m.}})\right),$$

where $A_{zz}^{(b)}$ and $C_{z,z}$ are the beam tensor analyzing power and spin correlation coefficient, respectively. The general expression for arbitrary orientation of the spins is more complicated and can be found in refs. 27-29. The main assumption used to estimate the 50% enhancement of the cross section for polarized DT fuel is that the reaction proceeds entirely through the $J^\pi = 3/2^+$ partial wave with an orbital relative angular momentum of the D+T pair $\ell = 0$ (that is, in an *s*-wave of relative motion). Under such an assumption, the unpolarized differential cross section is isotropic (that is, independent from the scattering angle). Furthermore, the integrals of $A_{zz}^{(b)}$ and $C_{z,z}$ over the scattering angle can be computed analytically and are 0 and $1/3$, respectively. This yields the estimate for the polarized reaction cross section $\sigma_{\text{pol}} \approx \sigma_{\text{unpol}}(1 + \frac{1}{2} p_z q_z)$, that is an enhancement factor of $\delta = 1.5$ when $p_z = q_z = 1$.

The study of the anisotropy in the unpolarized differential cross section stands as a first stringent test of this estimate. When investigating the angular differential cross section divided by the reaction cross section (its integral over the scattering angle), as done before in the experiment of ref. 30, these appear as deviations from unity. As shown in Fig. 2, a departure from a pure *s*-wave behavior is apparent. In particular, *p*-waves ($\ell = 1$) are responsible for the oblique slope, and *d*-waves ($\ell = 2$) for the making of a bump at 90°. Overall, we find good agreement with experiment once the centroid of the $3/2^+$ resonance is correctly located. It is worth noting that the degree of anisotropy does not exceed the 1.6% level, leading to an absolute variation of the differential cross section of about 6.6 mb between 0° and 180°. The overall good reproduction of the data gives once again evidence of the high-quality of the computed collision matrix. Thereafter we present our *ab initio* results including the phenomenological adjustment of the $3/2^+$ resonance centroid, and comment when appropriate, on its effect.

**Validation of model for polarized reaction observables.** As a further test, we computed $A_{zz}^{(b)}$ and $C_{z,z}$ from the components of the S-matrix using the formalism of the density matrix. As benchmark, we verified that (under the condition of an unpolarized target) we could reproduce the beam analyzing powers derived and computed independently. In principle both these observables can be measured in a laboratory by analyzing the differences with respect to the unpolarized cross section when the deuteron beam is vector- and tensor-polarized, the tritium is vector-polarized, and beam and target polarizations are aligned along the z-axis. In practice, however, only the tensor analyzing power at $\theta_{\text{c.m.}} = 0°$ has been measured in the energy region relevant for thermonuclear fusion ($0.24^{-0.9}_{+0.18}$ MeV)[31]. Our computed result ($-0.975$) agrees well with experiment ($-0.929 \pm 0.014$) (see also Supplementary Note 3). The only available experimental data to test the angular distribution of the differential cross section, and hence the contribution of partial waves other than the $J^\pi = 3/2^+, \ell = 0$ component at the relevant energies are measurements of the mirror D$^3$He fusion process. Such contribution of additional partial waves is exemplified in Fig. 3a, where we compare theoretical and experimental results for the D$^3$He tensor analyzing power at the deuteron incident energy of 0.424 MeV after subtraction of the s-wave contribution, which is simply given by the Legendre polynomial $-P_2(\cos\theta_{c.m.})$. Our results are in fair agreement with the experimental data[32], particularly for what concerns the shape of the distribution. At the same time, we find notable differences with respect to the predicted DT $A_{zz}^{(b)}$ at the corresponding energy of $E_D = 0.1$ MeV (where we take into account the difference in Q-values), highlighting a somewhat different partial-wave content in the two mirror reactions. This indicates that some caution has to be taken when using D$^3$He as a proxy for the study of polarization in the DT fusion process. All in all, Fig. 3a gives added confidence in the polarization observables predicted for the DT fusion. More details on the calculation of the D$^3$He reaction observables can be found in Supplementary Note 4.

**Reaction cross section enhancement.** The differential cross section for all angles is required for the computation of the polarized reaction cross section $\sigma_{pol}$ and the enhancement factor $\delta$, which we obtain (for any initial spin configuration) as the ratio of the latter to $\sigma_{unpol}$. As shown in Fig. 4a, at the deuteron incident energy of 100 keV the

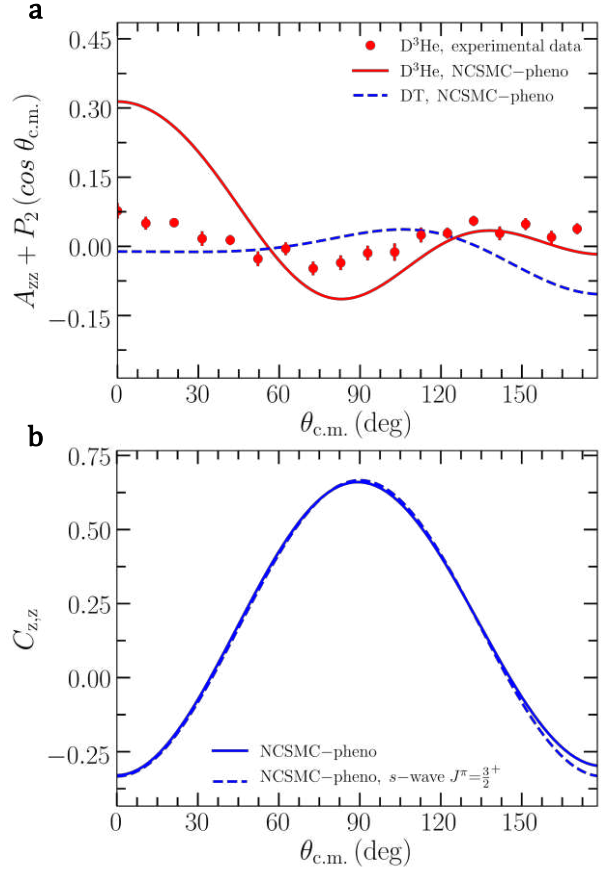

Fig. 3 | DT and D$^3$He polarization observables. **a** Computed and measured[32] tensor analyzing power ($A_{zz}^{(b)}$) of the D$^3$He fusion reaction as a function of the scattering angle in the center-of-mass (c.m.) frame, $\theta_{c.m.}$ at the deuterium incident energy of $E_D = 424$ keV compared to the results for the DT fusion reaction at $E_D = 100$ keV (the corresponding energy, once the difference in the Q-values of the two reactions is taken into consideration). The s-wave contribution to the tensor analyzing power, given by the Legendre polynomial $-P_2(\cos\theta_{c.m.})$, has been subtracted. The 'NCSMC-pheno' label stand for the results of the present calculations after a phenomenological correction of $-5$ keV to the position of the $3/2^+$ resonance. **b** Computed spin correlation coefficient ($C_{z,z}$) for the DT fusion at $E_D = 128$ keV. The results obtained by disregarding the contribution of partial waves beyond the $J^\pi = 3/2^+, \ell = 0$, labelled as 'NCSMC-pheno, s-wave $J^\pi = 3/2^+$', are also shown for comparison.

*ab initio* calculation recovers and confirms the ideal enhancement factor for $p_z, q_z = 1.0$, which is a result independent of the model space size and the phenomenological correction. Indeed, while our ab initio calculations show that the reaction is not exactly isotropic, at this energy the deviations of $A_{zz}^{(b)}$ and $C_{z,z}$ from a pure $J^\pi = 3/2^+, \ell = 0$ contribution are substantial only in the proximity of $\theta_{c.m.} = 180°$ (see, for example, Fig. 3b), and hence have only a minor effect on angle-averaged observables, such as the reaction cross section. We note that $\delta$ is nearly independent of the value of $p_{zz}$, indicating that the analyzing power of the deuterium does not play any role in the enhancement of the cross section (a consequence of the nearly-zero value of the integral of $A_{zz}^{(b)}$). However, the factor $\delta$ varies as a function of the energy and drops significantly above the deuteron incident energy of 0.8 MeV. This is shown in Fig. 4b for the maximum enhancement (which is found for $p_z q_z, p_{zz} = 1$). Interestingly, the peak value of the maximum enhancement (located around $E_D = 0.4$ MeV) is somewhat larger than the estimated 1.5 value. This is mainly an effect of $3/2^+$, $\ell = 2$ contributions. For comparison we also show the maximum enhancement obtained when we only include the $J^\pi = 3/2^+$, and $1/2^+$ partial waves with an orbital relative angular momentum of the D+T pair of $\ell = 0$. This shows the influence of $1/2^+$ components of the wave function below and above the $3/2^+$ resonance even in a purely s-wave picture of the reaction. When also the $1/2^+$ partial wave is removed, we recover the (energy independent) 1.5 estimate.

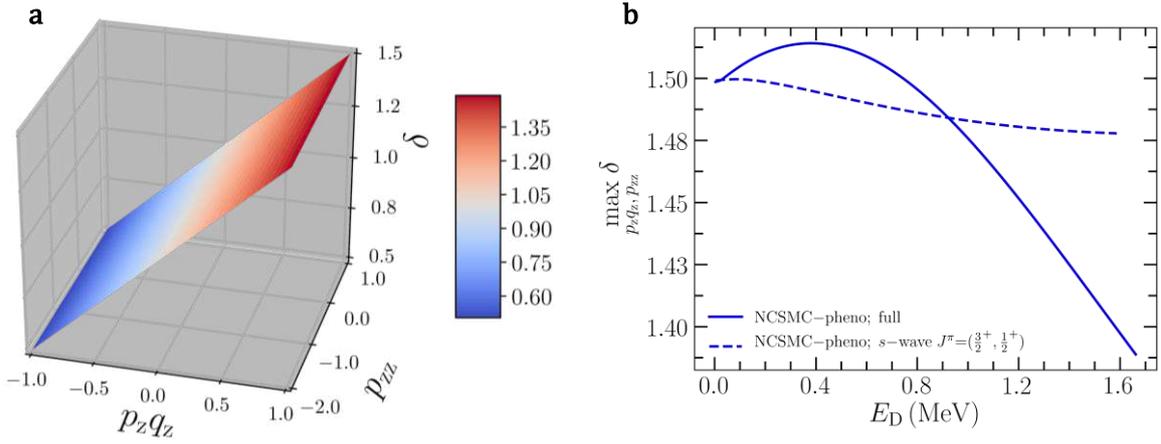

Fig. 4 | Enhancement factor of the polarized DT reaction cross section. a Present results for the enhancement factor ($\delta$) of the polarized DT reaction cross section at the deuterium incident energy of $E_D = 100$ keV as a function of the vector ($p_z q_z$) and tensor ($p_{zz}$) polarization of the deuterium and tritium. b Computed maximum enhancement factor (over all possible values of $p_z q_z$ and $p_{zz}$) of the polarized DT cross section as a function of the deuteron incident energy. The maximum enhancement is always found for $p_z q_z, p_{zz} = 1$. The 'NCSMC-pheno; full' label stand for the results of the present calculations including $\ell \neq 0$ partial waves after a phenomenological correction of $-5$ keV to the position of the $3/2^+$ resonance. Due to the energy scale of the figure, the enhancement factor obtained without such phenomenological correction (that is, the NCSMC result) is indistinguishable from the NCSMC-pheno curve. Also shown is the maximum enhancement factor obtained by retaining only the $\ell = 0$, $J^\pi = 3/2^+$ and $1/2^+$ partial waves, labeled as 'NCSMC-pheno; s-wave $J^\pi = (3/2^+, 1/2^+)$'.

**Reaction rate enhancement.** In Fig. 5 we show the polarized fusion reaction rate for typical values of vector and tensor polarization of the deuterium ($p_z, p_{zz}$) and tritium ($q_z$) that can be readily obtained in the laboratory, that is $p_z, p_{zz} = 0.8$ and $q_z = 0.8$, respectively. This quantity, obtained from averaging the reaction cross section over the distribution of the reactants' speeds (assumed to be Maxwellian)[33],

$$\langle \sigma v \rangle = \sqrt{\frac{8}{\pi \mu (k_B T)^3}} \int_0^\infty S(E) \exp\left(-\frac{E}{k_B T} - \sqrt{\frac{E_G}{E}}\right) dE,$$

is a measure of how rapidly the reaction occurs and is an important input in astrophysics and plasma simulations. The constant $\mu$ is the reduced mass of the reacting nuclei (D and T), $k_B$ and $T$ are respectively the Boltzmann constant and the temperature, $S(E)$ stands for the (computed) S-factor and $E_G$ is the Gamow energy given by $2\mu(\pi e^2)^2/\hbar^2$. In the figure we also compare our unpolarized reaction rate to those obtained from the parameterization of DT fusion data of Bosch and Hale[34], the phenomenological $R$-matrix fit of Descouvemont et al.[35], and the potential model calculation adopted in the NACRE compilation[36], which is intended for applications in astrophysics simulations. Overall, we find that they agree well even at energies above the resonance. In more detail, our calculation agrees best with the phenomenological $R$-matrix evaluation, particularly at higher energies where data are typically scarcer. In our case, the uncertainties due to the finiteness of the model space are indistinguishable from the line width. The convergence of our *ab initio* model is discussed in Supplementary Note 1. A further analysis of the systematic and statistical uncertainties associated with the adopted nuclear interaction model, such as those stemming from the order of the chiral expansion or the uncertainty in constraining its parameters, is presently computationally prohibitive (see also Supplementary Discussion). The phenomenological correction induces a global shift towards the reaction threshold, commensurate with that of the resonance centroid. In practice, this fine tuning is tightly constrained by the requirement to match S-factor data in the energy range below the resonant peak. The polarized reaction rate shows the same shape, albeit globally enhanced by a factor of ~1.32, in agreement with the approximate estimate for the chosen polarization. This result follows from the rather slow variation of the enhancement factor of the reaction cross section as a function of the energy in the narrow Gamow window (deuteron incident energies below a few hundred keV) where the product of the Maxwell-Boltzmann distribution with the tunneling probability of the nuclei through their Coulomb barrier is significantly different from zero. It is interesting to note that with polarization a reaction rate of equivalent magnitude as the apex of the unpolarized reaction rate is reached at lower temperatures, that is less than 30 keV compared to 65 keV (where both rates peak), as highlighted in Fig. 5 by the arrows. As a naive illustration, this means that by using polarized DT fuel the output of a standard fusion reactor could either be enhanced by 32% or its operational temperature decreased by as much as 45%. A more comprehensive discussion of the economics of using polarized fuel in the case of inertial confinement fusion can be found in ref. 4.

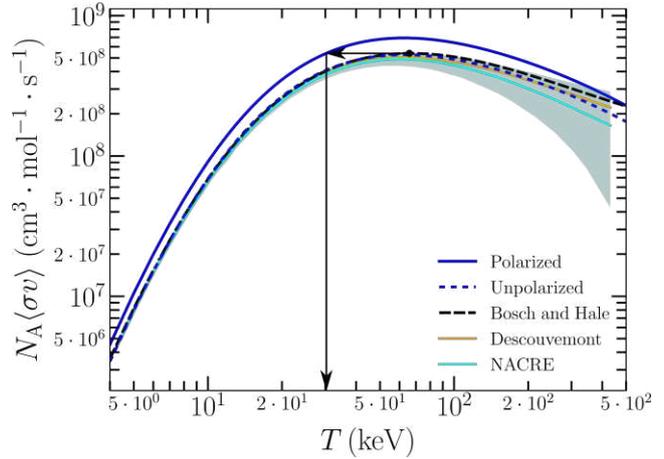

**Fig. 5 | DT reaction rate with and without polarization.** Comparison between the computed DT reaction rate ($N_A \langle \sigma v \rangle$, with $N_A$ the Avogadro number) for unpolarized and polarized fuel with aligned spins as a function of the temperature, $T$. The 'Polarized' and 'Unpolarized' labels stand for the present results obtained with the phenomenological correction of −5 keV to the position of the $3/2^+$ resonance (dubbed NCSMC-pheno). We use reactants' polarization parameters achievable in the laboratory, that is $p_z, p_{zz} = 0.8$ and $q_z = 0.8$. Also shown for comparison are the unpolarized reaction rates obtained from the widely adopted parametrization of the DT fusion cross section of Bosch and Hale[34] (labelled as 'Bosh and Hale'), from the $R$-matrix fit of Descouvemont[35] (labelled as 'Descouvemont') and from the NACRE compilation[36] (labelled as 'NACRE'). The arrows in the figure show that, with polarization, a reaction rate of equivalent magnitude as the apex of the unpolarized reaction rate is reached at lower temperatures.

**Angular distribution of the polarized reaction products.** While the deviations from a pure $J^\pi = 3/2^+, \ell = 0$ contribution are small and have only a minor effect, particularly on angle-averaged observables such as the reaction cross section or the reaction rate, they play a somewhat larger role on the angular distribution of the reaction products, especially when the reactants' spins are not in a parallel setting. In particular, while the tensor analyzing power $A_{zz}^{(b)}$ has virtually no impact on the enhancement factor, it is the main driver of the shape of the angular distribution of the polarized cross section, shown in Fig. 6. To better visualize the situation in the laboratory, in addition to the differential cross section in the c.m. frame we also plot the differential cross section in the laboratory frame as a function of the laboratory neutron and α-particle angles in yellow short-dashed and green dotted lines, respectively. The anisotropy of the angular differential cross section is highly sought after because it can be used to force the emitted neutrons and α to be two to five times more focused towards the reactor blanket (Fig. 6a and 6c), which collects the energy released, than along the polarization axis, or twice the exact opposite (see Fig. 6b). That is, the reaction products are more focused along the magnetic field. The former conditions can be achieved using only polarized deuterium or fully polarized DT fuel. The latter is obtained in the situation where the D and T spins are anti-aligned, leading to a reduction of the cross section of up to a factor of 0.5, as illustrated in Fig. 4a.

## Discussion

In conclusion, we have performed *ab initio* no-core shell model with continuum calculations with modern chiral EFT nucleon-nucleon and three-nucleon interactions for the DT fusion and its mirror D$^3$He reaction. We were able to reproduce the cross sections of these reactions with unpolarized reactants. Our calculations discriminate among DT reaction rates from phenomenological evaluations and demonstrate in detail the small contribution of $\ell > 0$ partial waves in the vicinity of the $3/2^+$ resonance. We predict the DT reaction rate for realistically polarized reactants ($p_z, q_z \sim 0.8$) and show that the reaction rate increases by about 32% compared to the unpolarized one and, further, the same reaction rate as the unpolarized one can be achieved at ~45% lower temperature. These results also endorse the application of the present approach to the evaluation of the polarized DD fusion, where the non-resonant character of the reaction prevents even a simple estimate of the enhancement factor in the ideal scenario of perfect spin-alignment of the reactants.

## Methods

**No-core shell model with continuum.** Our approach to the description of the DT fuson reaction is the *ab initio* no-core shell model with continuum (NCSMC) introduced in ref. 19 and applied to nucleon[9,11,12], deuterium[37], tritium and $^3$He induced reactions[38] and the three-cluster continuum dynamics of the Borromean $^6$He nucleus[39,40]. Presently, it is the only *ab initio* reaction method capable to efficiently describe complex light-nuclei reactions and in particular transfer reactions, though a complementary approach based on lattice effective field theory offers a more efficient avenue to the calculation of scattering and reactions induced by α particles[41].

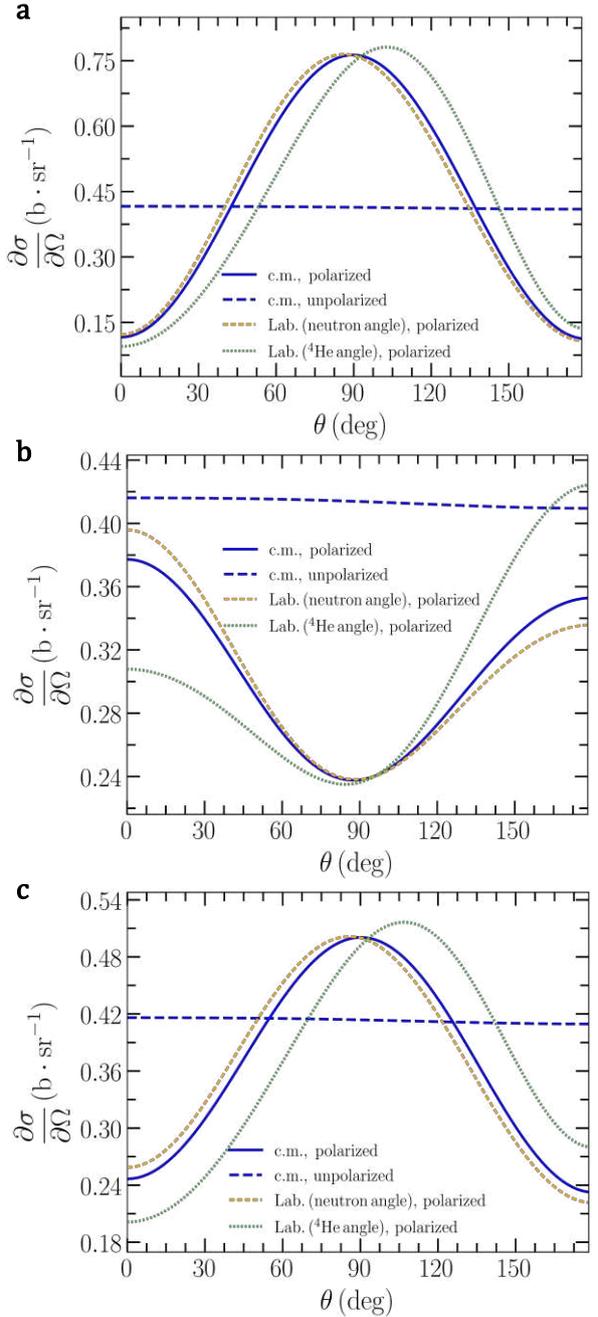

**Fig. 6 | Present results for the polarized DT differential cross section $\left(\frac{\partial \sigma}{\partial \Omega}\right)$.** Three polarization scenarios are shown: **a** With reactants' polarization parameters $p_z, p_{zz} = 0.8$, and $q_z = 0.8$; **b** with spins prepared in an antiparallel setting ($p_z = 0.8$, $q_z = -0.8$ and $p_{zz} = 0.8$); and **c** in the scenario in which only the deuterium is polarized ($p_z = 0.8$ and $p_{zz} = 0.8$). The incident deuterium energy is 100 keV. The 'c.m., polarized', 'Lab. (neutron angle), polarized', and 'Lab. ($^4$He angle), polarized' labels stand for the polarized differential cross section in the center-of-mass (c.m.) frame, and in the laboratory frame as a function of the neutron and $^4$He angles. Correspondingly, $\theta$ stands for the c.m., neutron and $^4$He angles. Also shown as a reference is the unpolarized cross section in the c.m. frame (labelled as 'c.m., unpolarized'). All results were obtained with the phenomenological correction of −5 keV to the position of the $3/2^+$ resonance (dubbed NCSMC-pheno).

The approach starts from the wave functions of each of the colliding nuclei and of the aggregate system, obtained within the *ab initio* no-core shell model (NCSM)[20] by working in a many-body harmonic oscillator (HO) basis. This is a configuration interaction method in which all nucleons are treated as active degrees of freedom and the model space includes all possible excitations of the system up to a maximum of $N_{max}$ quanta above the minimum-energy configuration. It then uses the NCSM static solutions for the aggregate system and continuous 'microscopic-cluster' states, made of pairs of nuclei in relative motion with respect to each other, as an over-complete basis to describe the full dynamical solution of the system. That is, the ansatz for the five-nucleon (A = 5) wave function takes the form of a generalized cluster expansion (here specifically shown for the present case of a $^5$He aggregate system):

$$|\Psi^{J^\pi I}\rangle = \sum_\lambda c_\lambda^{J^\pi I} |^5\text{He}; \lambda J^\pi I\rangle + \sum_\nu \int dr\, r^2\, \frac{\gamma_\nu^{J^\pi I}(r)}{r} \hat{A}_\nu \left|\Phi_{\nu r}^{J^\pi I}\right\rangle,$$

where $J, \pi$ and $I$ denote respectively total angular momentum, parity and isospin quantum numbers. The first term on the right-hand side of the equation is an expansion over the discrete energy-eigenstates of the $^5$He nucleus (indexed by λ) obtained within the NCSM method to incorporate the physics of the five nucleons in close contact. The second term is tailored to tackle the scattering and long-range clustering of the system. The index $\nu = \{\nu_{TD}, \nu_{\alpha n}\}$ runs over the reaction channels, defined by the mass partition (D+T and n+$^4$He, respectively) and the quantum numbers characterizing the reacting bodies and their relative motion. The continuum basis states $\Phi_{\nu r}^{J^\pi I}$ are antisymmetrized by the operator $\hat{A}_\nu$, and, in the case of the present binary collision, read:

$$\left|\Phi_{\nu_{TD}r}^{J^\pi I}\right\rangle = \left[[|^3\text{H}; \lambda_T J_T^{\pi_T} I_T\rangle|^2\text{H}; \lambda_D J_D^{\pi_D} I_D\rangle]^{S_{DT}I} Y_{\ell_{TD}}(\hat{r}_{TD})\right]^{J^\pi I} \frac{\delta(r-r_{TD})}{rr_{TD}},$$

and

$$\left|\Phi_{\nu_{\alpha n}r}^{J^\pi I}\right\rangle = \left[[|^4\text{He}; \lambda_\alpha J_\alpha^{\pi_\alpha} I_\alpha\rangle|n\rangle]^{S_{\alpha n}I} Y_{\ell_{\alpha n}}(\hat{r}_{\alpha n})\right]^{J^\pi I} \frac{\delta(r-r_{\alpha n})}{rr_{\alpha n}}.$$

The first set of continuum states describes the incoming T and D nuclei in relative motion, with $\vec{r}_{TD}$ the separation between their centers of mass, while the second set represents the outgoing wave of relative motion between the ejected α and neutron particles with separation $\vec{r}_{\alpha n}$. (Expressions in squared brackets denote angular momentum coupling.) The discrete coefficients $c_\lambda^{J^\pi I}$ and continuous amplitudes of relative motion $\gamma_\nu^{J^\pi I}(r)$ are obtained by solving the generalized eigenvalue problem derived from representing the non-relativistic Bloch-Schrödinger equation in the model space spanned by the discrete and continuum basis states of the NCSMC. The scattering matrix – and from it all reaction observables – are finally obtained by matching these solutions with the known asymptotic behavior of the wave function at $r = 18$ fm, using the coupled-channel $R$-matrix method on a Lagrange mesh[42,43].

**Details of the calculation.** We start from a five-nucleon Hamiltonian including NN[44] and 3N[45,46] interactions at the fourth and third order of chiral EFT, respectively, with a 500 MeV cutoff (also adopted in the studies of refs. 9 and 37). This interaction is then softened by the means of the similarity renormalization group (SRG) technique to a resolution scale of $\Lambda_{SRG} = 1.7$ fm$^{-1}$, enabling good convergence properties within the currently largest HO basis size achievable. The computational challenges of the present work limited such a basis size to a maximum number of HO excitations of $N_{max} = 11$. For the HO frequency, we chose the value of $\hbar\omega = 16$ MeV, which was found to speed up the convergence rate with respect to $N_{max}$ (see Supplementary Note 1).

Besides the size of the HO model space, the convergence properties of the present calculations are also affected by the number of discrete eigenstates of the $A$=2-, 3-, 4- and 5-nucleon systems used to construct the NCSMC trial wave function. We included the first fourteen discrete energy-eigenstates of the $^5$He system (two $J^\pi = 1/2^-$, three $3/2^-$, $5/2^-$, $7/2^-$, three $1/2^+$, two $3/2^+$, $5/2^+$, $7/2^+$), the ground state and up to 8 positive-energy eigenstates (5 in the $^3S_1$-$^3D_1$ and 3 in the $^3D_2$ channels) of the deuterium, and the ground states of the $^3$H and $^4$He nuclei. The close vicinity of the energy continuum of the deuterium, only bound by 2.224 MeV, leads to distortion effects the description of which necessitates the inclusion of positive-energy eigenstates[7,37]. Analogous distortion effects are less pronounced in the more bound triton and α particles, and are efficiently addressed indirectly through the inclusion of the eigenstates of the aggregate $^5$He system[9,10].

A particular challenge in the presence of 3N forces is the dependence on the parameter $E_{3max}$. This embodies the size of the three-nucleon single-particle HO basis used to represent the 3N interaction. For technical reasons, the largest $E_{3max}$ value computationally achievable is currently of 17 HO quanta. High energy 3N force components of the NCSMC Hamiltonian can be slowly converging as a function of this parameter. Since they represent a small perturbation with respect to the NN contribution, we omit them for basis states at the boundary of the model space.

**Phenomenological correction.** Remaining inaccuracies in the adopted chiral Hamiltonian prevent an accurate (of the order of less than a few keV) reproduction of the sub $p$-shell levels. This was already observed for the $^5$He system in, e.g., Figure 16 of ref. 10, which illustrates the residual imprecisions for the reproduction of $p$-shell spectroscopy. It is then not surprising that the DT fusion S-factor is not perfectly reproduced (see Fig. 1a). To address this difficulty, we treated the eigenvalue of the second $3/2^+$ NCSM energy-eigenstate (one of the static basis states that serve as input to represent our solution) as an adjustable parameter and constrained it to the value that yielded the best fit of the experimental S-factor data for energy below the resonance. In practice, this resulted in a shift of $-86$ keV of the $N_{max} = 11$ $^5$He $3/2^+$ eigenenergy computed within the NCSM, which was initially $-8.186$ MeV, while the microscopic n+$^4$He and D+T cluster potentials and all other characteristics of the scattering matrix continued to be predicted within the *ab initio* method. The amplitude of the correction is less substantial than it appears. In the NCSMC Hamiltonian, the coupling matrix elements between the aggregate system and microscopic-cluster states are given by the NCSM eigenvalues multiplied by the cluster form factor (the overlap between the two type of basis states). As a consequence, the effect of this adjustment is a considerably smaller shift of $-5$ keV of the resonance centroid $E_r$ extracted from the $3/2^+$ eigenphase shifts computed within the NCSMC, shown in Supplementary Table 1. Because the $3/2^+$ resonance is close to the D+T threshold, the S-factor is very sensitive to its centroid. Assuming a Breit-Wigner formula for the reaction cross section one can estimate the S-factor to be proportional to $1/E_r^2$ close to threshold and to follow a $1/E_{c.m.}^2$ slope after the resonance. This explains our results, and why our phenomenological adjustment is tightly constrained by reproducing the S-factor close to threshold. We refer to the modified calculation as NCSMC-pheno.

## Data Availability

The data acquired in this study are available from the corresponding author upon reasonable request.

## Acknowledgements

Computing support for this work came from the Lawrence Livermore National Laboratory (LLNL) institutional Computing Grand Challenge program. This article was prepared by LLNL under Contract DE-AC52-07NA27344. This material is based upon work supported by the U.S. Department of Energy, Office of Science, Office of Nuclear Physics, under Work Proposals No. SCW1158 and SCW0498, and by the Natural Sciences and Engineering Research Council of Canada (NSERC) Grants No. SAPIN-2016-00033. TRIUMF receives funding via a contribution through the Canadian National Research Council of Canada.

## Author Contributions

G.H. carried out the calculations in consultation with S.Q. and P.N. All authors discussed the results and contributed to the manuscript at all stages.

## Competing interests

The authors declare no competing financial and non-financial interests.

# Supplementary Information

*Ab initio* predictions for polarized deuterium-tritium thermonuclear fusion

Hupin et al.

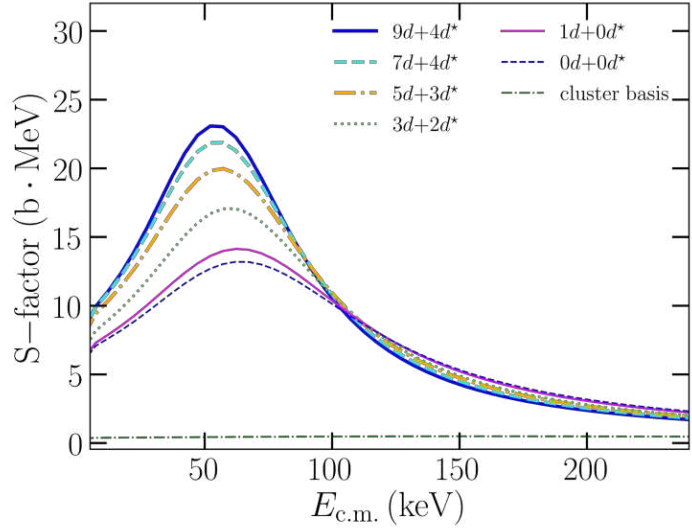

**Supplementary Figure 1** Convergence of the DT S-factor with increasing number of positive-energy eigenstates of the deuterium. Results obtained within the NCSMC approach at the harmonic oscillator model-space size $N_{max} = 11$ plotted as a function of the energy in the center-of-mass frame, $E_{c.m.}$. The '$nd+md^*$' labels stand for the S-factors obtained by including the first $n$ positive-energy eigenstates in the $^3S_1$-$^3D_1$ channel plus the first $m$ eigenstates in the $^3D_2$ channel of $^2$H. Results obtained within the cluster basis alone (with the deuterium ground state only) are shown as reference and are labeled as 'cluster basis'.

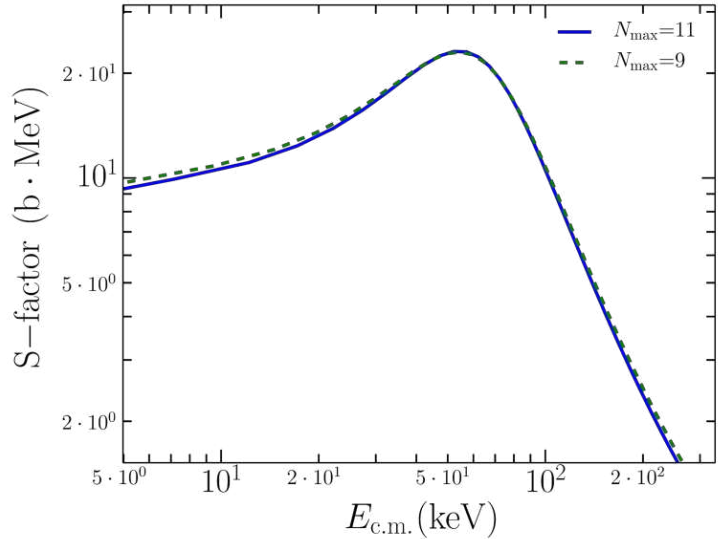

**Supplementary Figure 2** Convergence of the DT S-factor with increasing harmonic-oscillator model-space size. The labels '$N_{max} = 9$' and '$N_{max} = 11$' stand for the DT S-factors obtained within the NCSMC approach at $N_{max} = 9$, and 11 (currently the largest achievable). In the figure, $E_{c.m.}$ is the energy in the center-of-mass frame.

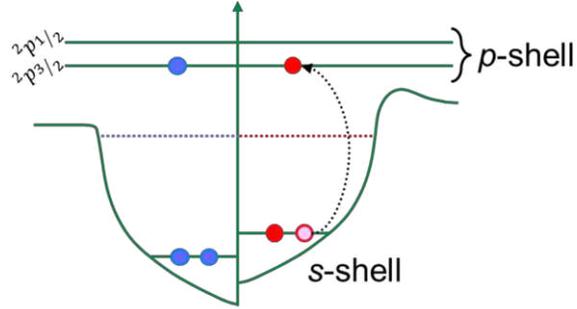

**Supplementary Figure 3** Simple sketch of the structure nature of the DT $3/2^+$ resonance. Neutrons are in blue and protons in red.

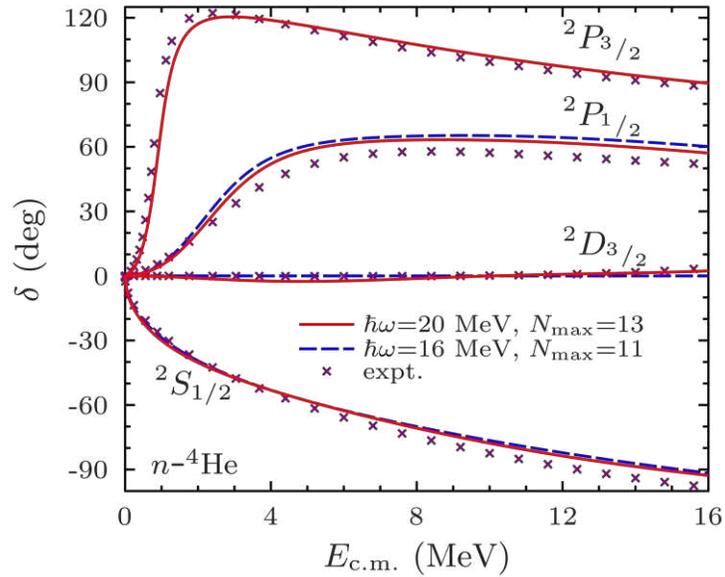

**Supplementary Figure 4** Neutron-$^4$He phase shifts below the fusion reaction threshold. Results in the $^2S_{1/2}$, $^2P_{1/2}$, $^2P_{3/2}$, and $^2S_{3/2}$ partial waves obtained in the NCSMC model space without D+T cluster states. The labels '$\hbar\omega = 20$ MeV, $N_{max} = 13$' and '$\hbar\omega = 16$ MeV, $N_{max} = 11$' stand, respectively, for the phase shifts obtained using the $\hbar\omega = 20$ MeV, $\Lambda_{SRG} = 2.0$ fm$^{-1}$, $N_{max} = 13$ and $\hbar\omega = 16$ MeV, $\Lambda_{SRG} = 1.7$ fm$^{-1}$, $N_{max} = 11$ sets of NCSMC parameters. An accurate $R$-matrix parametrization of experimental data (G. M. Hale, personal communication[1]) is shown as a reference, labelled as 'expt.'.

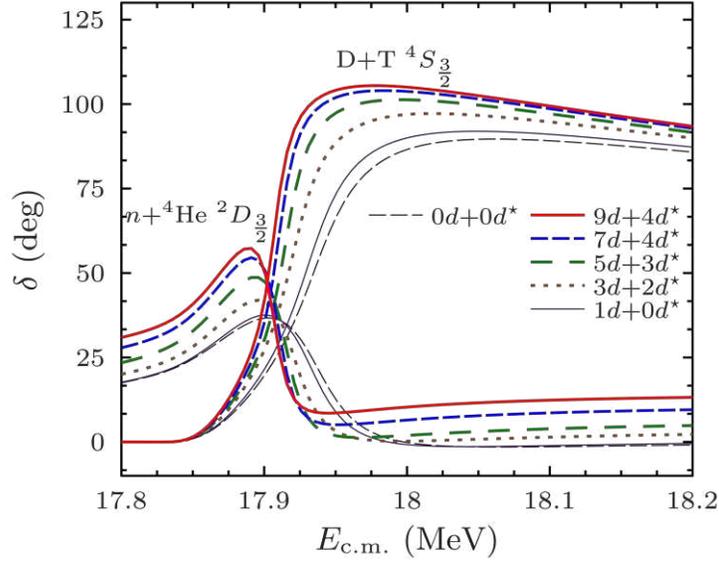

**Supplementary Figure 5** Diagonal phase shifts in the entrance and exit channels of the DT fusion. Convergence of the (real part of the) D+T $^4S_{3/2}$ and n+$^4$He $^2D_{3/2}$ diagonal phase shifts (characterizing, respectively, the entrance and exit scattering states of the DT fusion reaction) in the $J^\pi = 3/2^+$ channel with increasing number of positive-energy eigenstates of the deuterium, as obtained within the NCSMC approach. The '$nd+md^*$' labels stand for the phase shifts obtained by including the first $n$ positive-energy eigenstates in the $^3S_1$-$^3D_1$ channel plus the first $m$ eigenstates in the $^3D_2$ channel of $^2$H.

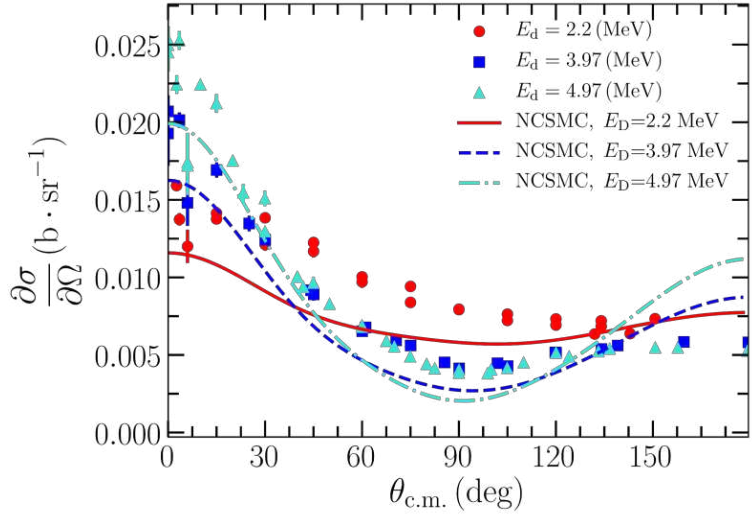

**Supplementary Figure 6** Unpolarized differential cross section. Comparison of computed and measured[2-6] differential cross sections $\left(\frac{\partial \sigma}{\partial \Omega}\right)$ in the center-of-mass (c.m.) frame at the deuterium incident energies of $E_D = 2.2, 3.97$ and $E_D = 4.97$ MeV, as a function of the scattering angle in the c.m. frame, $\theta_{c.m.}$. The labels 'NCSMC' and 'Expt.' stand, respectively for the present results and the experimental data.

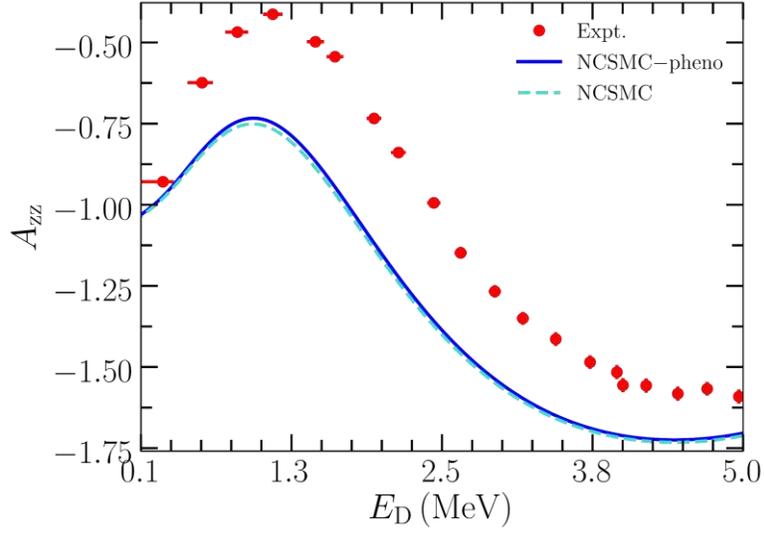

**Supplementary Figure 7** Tensor analyzing power for the DT reaction. Comparison between present results and measurements[7] for the tensor analyzing power ($A_{zz}^{(b)}$) at energies below the breakup threshold and center-of-mass angle $\theta_{c.m.} = 0°$ as a function of the deuterium incident energies of $E_D$. The labels 'Expt.', 'NCSMC', and 'NCSMC-pheno' stand, respectively, for the experimental data, the present calculation, and the results of the present calculation after a phenomenological correction of $-5$ keV to the position of the $3/2^+$ resonance.

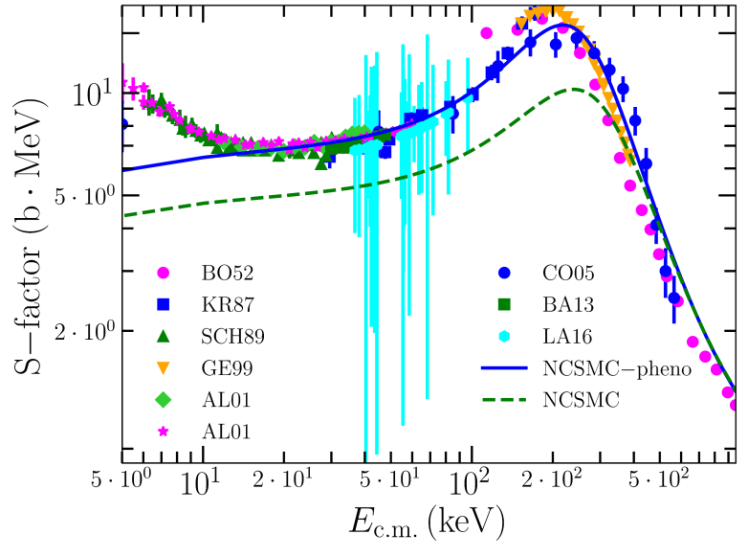

**Supplementary Figure 8** D³He astrophysical S-factor. The results of the present calculation before ('NCSMC') and after ('NCSMC-pheno') phenomenological adjustment of the $3/2^+$ resonance are compared with the experimental data of refs. 8-15 (labelled, in order, 'BO52', 'KR87', 'SCH89', 'GE99', 'AL01', 'CO05', 'BA13', and 'LA16'). In the calculations, the harmonic oscillator model space size is limited to $N_{max} = 9$ for computational reasons. $E_{c.m.}$ denotes the energy in the center-of-mass frame.

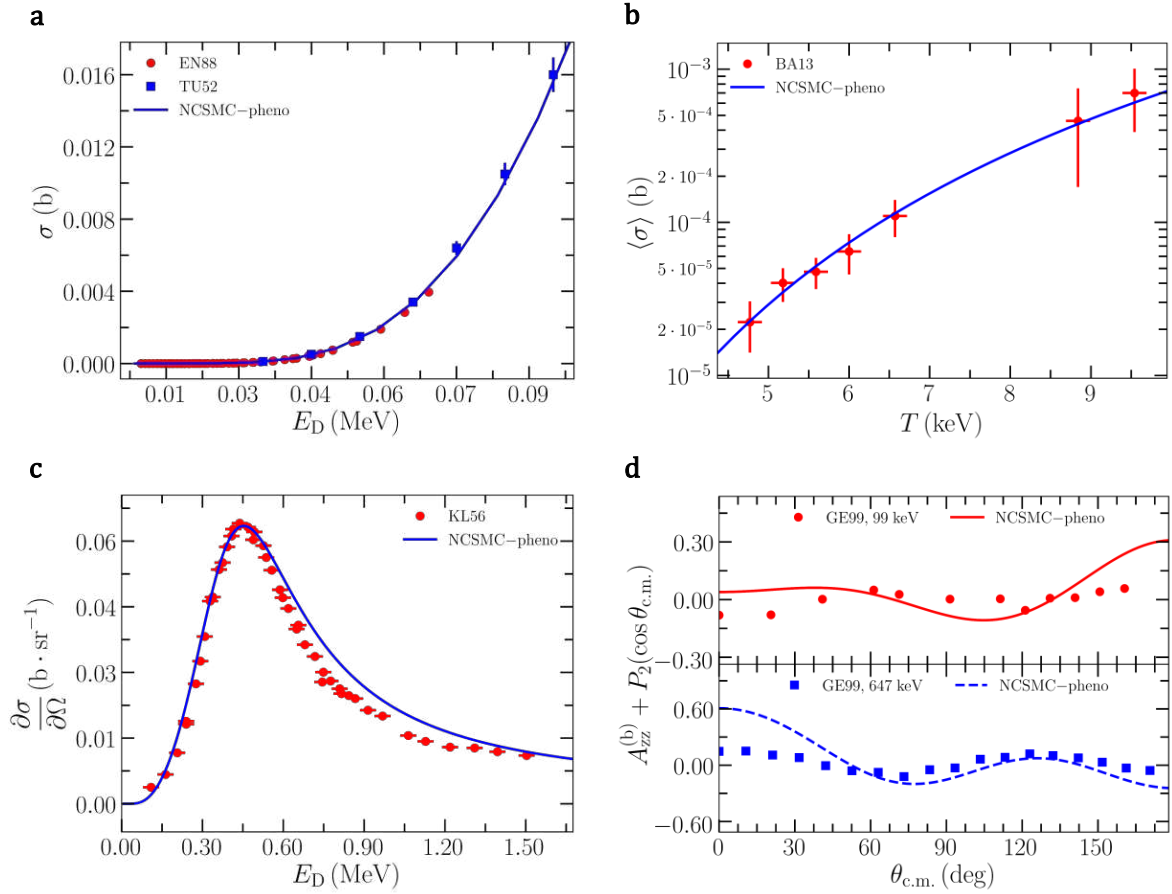

**Supplementary Figure 9** D³He reaction observables. **a** Fusion cross section ($\sigma_{\text{tot}}$) in the energy range below the resonant peak compared to the measurements of refs. 16 ('TU52') and 17 ('EN88'). **b** Computed temperature averaged cross section ($\langle\sigma\rangle$) compared to the data of ref. 14 ('BA13'). **c** Computed differential cross section $\left(\frac{\partial\sigma}{\partial\Omega}\right)$ at the center-of-mass (c.m.) scattering angle of $\theta_{\text{c.m.}} = 90°$ as a function of the impinging deuterium energy ($E_D$) compared to the experimental data of ref. 18 ('KL56'). **d** Computed $A_{zz}^{(b)}$ tensor analyzing power after subtraction of the $J^\pi = 3/2^+$, $\ell = 0$ contribution [given by the Legendre polynomial $-P_2(\cos\theta_{\text{c.m.}})$] compared to the data from ref. 11 ('GE99') at the D incident energies of $E_D = 99$ keV (top) and 641 keV (bottom). In the figures 'NCSMC-pheno' stands for the results of the present calculations after phenomenological adjustment of the $3/2^+$ resonance.

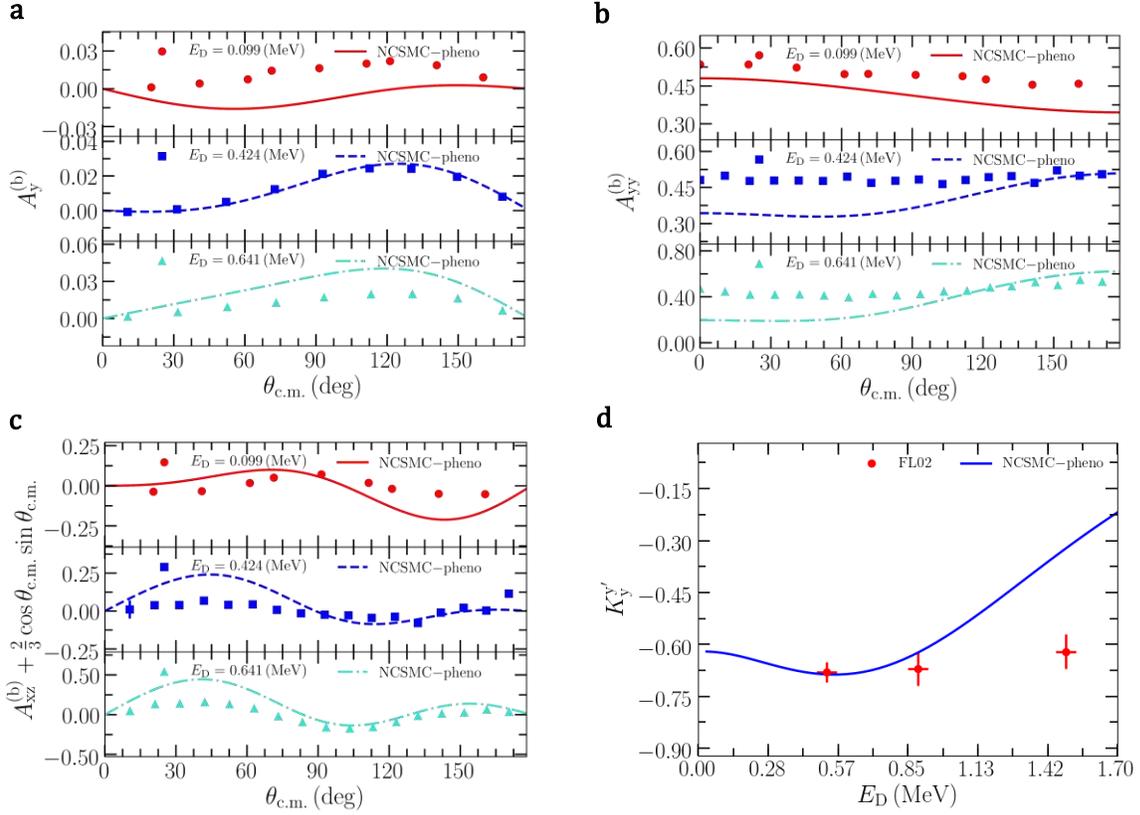

**Supplementary Figure 10** D³He polarization observables. **a-c** Computed $A_y^{(b)}$ vector and $A_{yy}^{(b)}$ and $A_{xz}^{(b)}$ tensor analyzing powers compared to the data from ref. 11 (circles, squares, and triangles) at the D incident energies of $E_D = 99$ keV (top), $E_D = 424$ keV (middle), and $E_D = 641$ keV (bottom). The $J^\pi = 3/2^+$, $\ell = 0$ contribution (given by the Legendre polynomial $-\frac{2}{3}\cos\theta_{c.m.}\sin\theta_{c.m.}$) is subtracted in panel **c**. **d** Computed polarization transfer coefficient ($K_y^{y'}$) around the reaction threshold at the center-of mass angle of $\theta_{c.m.} = 0°$ compared to the experimental data of ref. 19 (circles). In the figures 'NCSMC-pheno' stands for the results of the present calculations after phenomenological adjustment of the $3/2^+$ resonance.

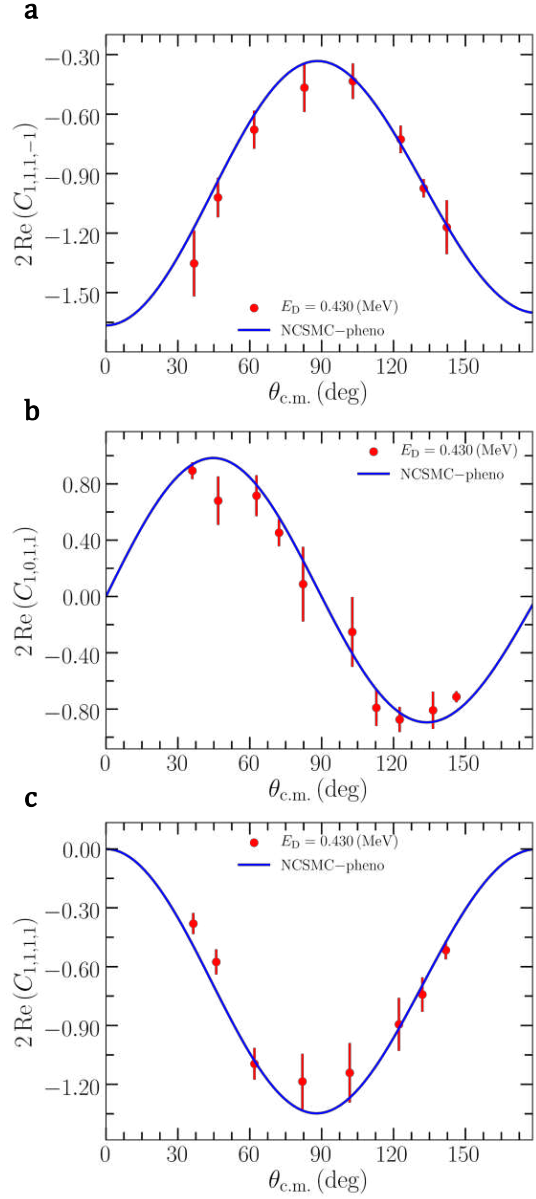

**Supplementary Figure 11** D³He spin correlation coefficients. Computed $C_{1,1,1,-1}$ (**a**), $C_{1,0,1,1}$ (**b**) and $C_{1,1,1,1}$ (**c**) spherical spin correlation coefficients as a function of the center-of mass angle, $\theta_{c.m.}$, compared to the experimental data from ref. 20 (circles) at the D incident energies of $E_D$ = 430 keV. In the figures 'NCSMC-pheno' stands for the results of the present calculations after phenomenological adjustment of the $3/2^+$ resonance.

**Supplementary Table 1** $^5$He $3/2^+$ resonance. Centroid and width of the $3/2^+$ resonance derived from the DT eigenphase shifts computed within the cluster basis alone ('Cluster basis'), and the full NCSMC basis before ('NCSMC') and after ('NCSMC-pheno') a phenomenological correction of $-5$ keV to the position of the $3/2^+$ resonance. The centroid is defined as the energy for which the first derivative of the eigenphase shifts is maximal, while the width is computed as twice the inverse of the derivative of the eigenphase shift at the resonance energy. Values derived from an $R$-matrix analysis of experimental data are shown as reference[21].

| $^5$He$(3/2^+)$ | Cluster basis (D g.s. only) | Cluster basis | NCSMC (D g.s. only) | NCSMC | NCSMC-pheno | $R$-matrix |
|---|---|---|---|---|---|---|
| $E_r$ (keV) | 105 | 120 | 65 | 55 | 50 | 47 |
| $\Gamma_r$ (keV) | 1100 | 570 | 160 | 110 | 98 | 74 |

**Supplementary Table 2** Convergence of the $^5$He $(3/2^+)$ eigenenergy. Relative difference with respect to the extrapolated infinite model space result of the eigenvalue of the $^5$He $(3/2^+)$ resonance computed within the NCSM approach as function of the harmonic oscillator (HO) basis size $N_{max}$. The notation '$\hbar\omega$' and '$\Lambda_{SRG}$' denote, respectively, the HO frequency and resolution scale of the similarity renormalization group transformation adopted in the two sets of displayed calculations.

| $N_{max}$ | $\hbar\omega=20$ MeV, $\Lambda_{SRG}=2.0$ fm$^{-1}$ | $\hbar\omega=16$ MeV, $\Lambda_{SRG}=1.7$ fm$^{-1}$ |
|---|---|---|
| 7 | 78.70% | 42.29% |
| 9 | 45.04% | 18.85% |
| 11 | 25.68% | 8.41% |
| 13 | 13.78% | - |

## Supplementary Note 1: Convergence of the calculation

Achieving convergence with respect to the number of eigenstates of the aggregate $^5$He system is straightforward. All eigenstates in a large range of energies around the region of interest for the deuterium-tritium (DT) fusion can be effortlessly included. The second $3/2^+$ eigenstate is exceptionally impactful. This can be clearly seen in Supplementary Figure 1 by comparing the S-factor (dominated by the $3/2^+$ component of the wave function) computed within the full no-core shell model with continuum (NCSMC) model space with the results obtained within the cluster basis alone. Evidently, the configuration where all five nucleons are in close contact plays an essential role.

Supplementary Figure 1 also shows the somewhat slow but steady convergence pattern of the S-factor with respect to the number (in order of increasing energy) of positive-energy eigenstates of the D projectile included in addition to the (negative-energy) ground state. We used the notation $d$ and $d^*$ for the eigenstates in the $^3S_1$-$^3D_1$ and $^3D_2$ channels of $^2$H, respectively. Since the number of available positive-energy eigenstates of the D projectile depends on the harmonic oscillator (HO) basis size $N_{max}$ (higher $N_{max}$ means more states to discretize the $^2$H continuum), it is instructive to compare this figure with the convergence in $N_{max}$ (Supplementary Figure 2). The agreement between the $N_{max} = 9$ and 11 calculations (both including the maximum number of available deuteron states) is quite good.

It should be noted that the S-factor is extremely sensitive to changes in the position of the $3/2^+$ resonance, acting as a magnifying glass. For example, the 35 keV shift to lower energies in the resonance position between the results obtained within the cluster basis alone and the full NCMC model space (see Supplementary Table 1) results in about one order of magnitude increase of the S-factor amplitude. In comparison, the positive-energy eigenstates of the D projectile contribute a shift of 10 keV (15 keV) in the $3/2^+$ resonance centroid when the eigenstates of the aggregate $^5$He system are (are not) included in the model space. This suggests that there is a strong similarity between some cluster basis states built from the discretization of the D energy continuum and some static solutions of the $^5$He aggregate. In this contest, it becomes also clear that the fine-tuning of the $3/2^+$ resonance centroid required to accurately reproduce the experimental S-factor is extremely challenging, given remaining inaccuracies of the adopted chiral interactions for p-shell nuclei. For this reason, we opted for a phenomenological fine tuning as explained in the Methods section.

The $3/2^+$ resonance, which drives the massive enhancement of the S-factor, schematically consists in a proton promoted from the s-shell onto the $^2p_{3/2}$ sub-shell, as shown in the sketch of Supplementary Figure 3. The energy required is of the order of the splitting between major HO s- and p-shells and can be inferred from the experimental spectra as approximatively the excitation energy of the $3/2^+$ resonance, that is 16.84 MeV. This explains why the HO frequency of $\hbar\omega = 16$ MeV chosen in this work contributes to speeding up the convergence of our calculations. Additionally, to ensure the convergence of our calculation within the computationally achievable largest model space ($N_{max} = 11$), we used a similarity renormalization group (SRG) resolution scale of $\Lambda_{SRG} = 1.7$ fm$^{-1}$. To further analyze our parameters' choice, in Supplementary Figure 4 we compare the present n + $^4$He elastic scattering phase shifts (obtained in the NCSMC model space without D+T cluster states) with those previously obtained for $\hbar\omega = 20$ MeV, $\Lambda_{SRG} = 2.0$ fm$^{-1}$ and $N_{max} = 13$[22,23]. We can see that the agreement is excellent. The fast convergence rate achieved within the present choice of parameters is even more manifest when analyzing the dependence on $N_{max}$ of the $^5$He $3/2^+$ eigenenergy computed within the NCSM. This is illustrated in Supplementary Table 2, where we show the difference between the computed eigenenergy at a given $N_{max}$ relative to the extrapolated energy at $N_{max} \to \infty$. The frequency closer to the major HO shell splitting performs much better. At $N_{max} = 11$ the relative difference goes down from 25.68% at $\hbar\omega = 20$ MeV, $\Lambda_{SRG} = 2.0$ fm$^{-1}$ to 8.41% at $\hbar\omega = 16$ MeV, $\Lambda_{SRG} = 1.7$ fm$^{-1}$, more than a factor of two.

## Supplementary Note 2: Reaction Mechanism

The centroid and width of the $3/2^+$ resonance computed within the NCSMC-pheno (i.e., by applying the phenomenological correction of $-5$ keV to the position of the $3/2^+$ resonance discussed in the Methods) are in good agreement with those extracted from the $R$-matrix analysis of data of ref. 21, particularly considering that the latter values were obtained from the $S$-matrix pole rather than from the eigenphase shift, as done here. In addition, the magnitude and width of the computed total n-$^4$He cross section in the energy region of the $3/2^+$ resonance are in a good agreement with the measurements of Haesner et al.[24], though the position of its peak is overestimated by about 1%. This is due in part to a slight difference between computed (17.8 MeV) and experimental (17.6 MeV) Q-value, and more in general to remaining inaccuracies of the adopted chiral interactions for $p$-shell nuclei which cannot be entirely corrected using the minimal (single-parameter) phenomenological adjustment adopted in this work. In Supplementary Figure 5, we show the (real part of the) phase shifts extracted from the diagonal elements of the $S$-matrix in the $J^\pi = 3/2^+$ channel, namely the $d$-$^3$H $^4S_{3/2}$ and $n$-$^4$He $^2D_{3/2}$ partial waves belonging, respectively, to the entrance and exit channels of the reaction. Similar to our more limited work of ref. 25, we find a sharp resonant behavior in the $d$-$^3$H $^4S_{3/2}$ partial wave, but the $n$-$^4$He $^2D_{3/2}$ phase shift is broader and does not cross 90°. Supplementary Figure 5 also highlights the influence of the deuterium continuum. We would like to stress that this reaction mechanism highlights the fundamental role played by the tensor force, present in both nucleon-nucleon (NN) and three-nucleon (3N) components of the nuclear interactions. In addition, the important role of the 3N force in reproducing the position of the resonance centroids and the splitting of the $^2p_{3/2}$ and $^2p_{1/2}$ sub $p$-shell levels has also become evident in the last decade[23,26]. Because of this (and the fact that the SRG procedure we use to accelerate the convergence generates induced 3N forces) the inclusion of 3N forces was essential to achieving the present accurate results for the DT fusion.

## Supplementary Note 3: Comparison to higher energy data

Our approach is presently valid up to the threshold of the dissociation of the deuterium projectile or 2.224 MeV. Above such energy, it represents an approximation. Nevertheless, in Supplementary Figures 6 and 7 we present a comparison with higher-energy data for the angular differential cross section and tensor analyzing power, respectively. At the deuteron energy of 2.2 MeV and above, the *ab initio* angular differential cross section systematically underestimates the data[2-6], though the shape of the angular distribution is qualitatively reproduced. The tensor analyzing power (presented in Supplementary Figure 7), which is inversely proportional to the differential cross section, further magnifies the difference between theory and experiment. At the origin of this discrepancy are higher-energy $^5$He resonances (known experimentally and in evaluations[27]) that come into play a few MeV above the peak-energy of the DT fusion, due to the population of the nearby $^2p_{1/2}$ subshell (see the sketch of Supplementary Figure 3). Lacking an exact treatment of three-cluster dynamics and given remaining inaccuracies of the adopted chiral Hamiltonian in reproducing the $p$-subshell ordering that affects the underlying phase shifts, these resonances are not reproduced with the required level of accuracy (within ~10 keV). We note that this only affects the cross section at higher energies. In particular, around the energy of interest for fusion applications (~100 keV), the tensor analyzing power is in good agreement with experiment.

## Supplementary Note 4: Calculation for the mirror D³He reaction

In Supplementary Figure 8, we compare our computed D³He S-factor and its phenomenological correction to available data[8-15]. All parameters of the NCSMC calculation match those of the used for the DT reaction (see Supplementary Note 1) but the $N_{\max}$ value, which in this case is limited to nine major shells for computational reasons. As in the DT case, the centroid position of the $3/2^+$ resonance of ⁵Li is overestimated and needs to be corrected phenomenologically. Once again, the adjustment of the $3/2^+$ resonance is strongly constrained by reproducing the S-factor from ~20 keV to energies below the resonance. At lower energies (below ~20 keV), the prediction is expected to disagree with data due to laboratory electron screening effects, which enhance the cross section masking the ("bare") nuclear S-factor. At the peak of the S-factor, the experimental picture is somewhat uncertain. Our results are in good agreement with the data of ref. 13 ('CO05'). The computed peak value of the reaction cross section (798 mb) is in good agreement with the experimental value of $777 \pm 33$ mb reported by Geist et al.[11] ('GE99'). However, the position of the peak is found at 450 keV, 24 keV above the energy reported in ref. 11. This slight energy shift is at the origin of the discrepancy between our calculation and the S-factor of Geist et al. In the present (restricted) $N_{\max} = 9$ model space, a 426 keV peak energy is inconsistent with the behavior of the S-factor at lower energy. There, we find good agreement with the total reaction cross section and thermalized cross section data of refs. 16, 17 and 14, respectively, owing to the tight constraint imposed by this energy regime on our phenomenological adjustment (see Supplementary Figures 9a and 9b). In Supplementary Figure 9c we compare the computed differential cross section at $\theta_{c.m.} = 90°$ with the experimental data up to $E_D = 1.6$ MeV of Klucharev et al.[18]. Overall, a small overestimation of data is noticeable above the fusion peak suggesting once again that the width of the $3/2^+$ resonance may be slightly overestimated. Based on the trend shown by the DT results of Supplementary Figure 2, we expect that an $N_{\max} = 11$ calculation would yield a narrower cross section peak in closer agreement with the experimental data of both Geist et al and Klucharev et al. We also computed an array of polarization observables, namely the vector $A_y^{(b)}$ and tensor $A_{zz}^{(b)}, A_{yy}^{(b)}$, and $A_{xz}^{(b)}$ analyzing powers at the deuteron incident energies of 99, 424 and 641 keV, and the polarization transfer coefficient $K_y^{y'}$ at $\theta_{c.m.} = 0°$, to compare to the measurements reported in refs. 11 and 19, respectively. The comparisons for the analyzing powers are shown in Fig. 3a of the Results section, Supplementary Figures 9d, and 10a-c, while that for the polarization transfer coefficient is presented in Supplementary Figure 10d. In general, we find fairly good agreement with the experimental data at 424 keV, while at the left and right of the peak we tend to obtain a good description of the overall angular and energy dependence but somewhat overestimate the amplitude. This can be traced back to the modest overestimation of the reaction cross section. Finally, in Supplementary Figure 11 we compare our computed spin correlation coefficients $C_{1,1,1,-1}$, $C_{1,0,1,1}$ and $C_{1,1,1,1}$ to the experimental data of ref. 20, at the incident deuteron energy of 430 keV. This is the only existing spin-correlation experiment in the resonance region for this reaction, and the most significant and direct test of our calculations for the polarized fusion. In our notation, the coefficients are written as spherical tensors with the initial (final) two indices corresponding to the rank and projection of the tensor moments of the beam (target). Our calculation agrees well with the experimental data. This stands as a chief validation of our predictions for the polarized DT fusion. Overall, our *ab initio* method together with modern chiral NN+3N interactions are able to reproduce both the DT fusion and its mirror D$^3$He reaction. This is a major step forward compared to the results obtained in our earlier work[25].

## Supplementary Discussion: Uncertainties of the calculation

In the present work, uncertainties derive either from the many-body model used to solve the five-body Schrödinger equation or from the employed nuclear Hamiltonian. The formers are addressed in Supplementary Note 1. There, we show that our calculation is converged with respect to the three parameters $\hbar\omega$, $N_{max}$ and $\Lambda_{SRG}$. Thus, uncertainties from the many-body technique are particularly small, typically close to the size of the line width as exemplified in Fig. 6 of the Results section. To demonstrate that the accuracy of the present application is not accidental, we computed the D$^3$He mirror reaction and compared both unpolarized and polarized reaction observables to data. We obtained satisfactory agreement with data that further validates our predictions for the DT polarized observables. On the other hand, it is computationally extremely challenging for the time being to give an estimate of the uncertainties pertaining the nuclear Hamiltonian. We use a chiral EFT Hamiltonian that has been proven to reproduce properties of the $A = 3, 4, 5, 6$ nuclei, including *p*-shell physics. Based on the fact that other chiral EFT Hamiltonians have emerged that fail to reproduce the low-lying *p*-waves of the $^5$He system, it is expected that uncertainties from the nuclear interaction model may be significant, but this remains to be investigated.

## Supplementary References